\documentclass[prl,preprint]{revtex4}
\usepackage{graphicx}
\usepackage{dcolumn}
\usepackage{bm}

\newcommand{\be}{\begin{equation}}
\newcommand{\ee}{\end{equation}}
\newcommand{\bea}{\begin{eqnarray}}
\newcommand{\eea}{\end{eqnarray}}

\begin{document}

\preprint{}
\title{Condensed Multiwalled Carbon Nanotubes as Super Fibers}
\author{Zhiping Xu}
\author{Lifeng Wang}
\author{Quanshui Zheng}
\email{zhengqs@tsinghua.edu.cn}
\affiliation{Department of Engineering Mechanics, Tsinghua University, Beijing 100084,
China}

\begin{abstract}
The ultra-low intershell shear strength in carbon nanotubes (CNTs) has been
the primary obstacle to applications of CNTs as mechanical reinforcements.
In this paper we propose a new CNT-system composed of comprising of coaxial
cylindrical shells of sp$^{2}$-bonded carbons with condensed intershell
spacings. Our atomistic calculations show that such condensed multiwalled
carbon nanotubes (CMWNTs) can greatly enhance intershell shear strengths by
several orders, and can simultaneously generate higher tensile strengths and
moduli respectively than those of ordinary CNTs. It has further shown that
CMWNTs can maintain thermally stable up to 2,000 K. By taking advantage of
the primary enhancement mechanism of CMWNTs, a method of producing CMWNTs is
therefore proposed tentatively. It is believed that CMWNTs featured with
those properties can be taken as excellent candidates of super fibers for
creating space elevators.
\end{abstract}

\maketitle

It may sound a crazy idea to build up a space elevator for delivering
payloads from the Earth into space, the ultimately extra-low delivering
cost, possibly US\$10 per kilogram, makes space elevator as the only
technology for exploiting solar power in a big way that could significantly
brighten the world's dimming energy outlook \cite{05-Edwards}. To meet the
biggest challenges of building a space elevator, the key issue lies in
creating a superstrong, lightweight cable that can stretch over 36,000 km
between the Earth and a geostationary space station. This requires at the
minimum a specific tensile strength (namely the tensile strength-specific
gravity ratio) of $48.5\,\mathrm{GPa}$ (for the graphite specific gravity
2.25), much higher than those of all natural or\_artifical materials, e.g. $%
1.36\,\mathrm{GPa}$ for graphite whiskers, or $0.08\,\mathrm{GPa}$ for high
carbon steel wires, before the discovery of carbon nanotubes (CNTs) \cite%
{91-Iijima}. It is known that sp$^{2}$ or carbon-carbon bond is the
strongest bond in the nature. A single-walled carbon nanotube (SWNT) can be
viewed as a cylindrical shell of sp$^{2}$-bonded monolayered carbon atoms,
and a multi-walled carbon nanotube (MWNT) an assembly of SWNT-like shells
with interwall spacings $\sim 0.34\,\mathrm{nm}$ in average. Both
experimental investigations \cite{96-Treacy,00-Yu} and atomic simulations 
\cite{92-Robertson,96-Yakobson,98-Hernandez} using molecular dynamics or
first-principals indicate that CNTs have a tensile Young's modulus of $\sim
1\,\mathrm{TPa}$ (10$^{11}\,\mathrm{Pa}$), and atomic simulations further
show that ideal or defect-free CNTs can have a tensile failure strain as
high as $16$ to $24\%$ \cite{96-Yakobson,02-Belytschko,03-Dumitrica}. These
would give the pristine CNTs a specific tensile strength of $70$ $%
\allowbreak $to $\allowbreak 100\,\mathrm{GPa}$, which can meet the
requirement of space elevator. Wide discrepancy, however, exists between the
theoretical and experimental results of CNT's failure strain and tensile
strength. The most extensive fracture measurements of MWNTs \cite{00-Yu}
have shown that failure strains are ranged from $2$ to $13\%$ with the MWNTs
breaking in the outermost wall. Quantum mechanical calculations \cite%
{04-Mielke} have explained the markedly reduced failure strain with vacancy
defects and, in particular, large holes. And to understand the fractional
breaking, Yu \textit{et al.} \cite{00-Yu} have found that the intershell
shear strength is ultra-low and predominantly originated from the van der
Waals interactions, with a value of $0.3\,\mathrm{MPa}$ comparable to the
shear strengths ($0.25$ to $0.75\,\mathrm{MPa}$) of high quality crystalline
graphite. Similar ultra-low intershell shear strength ($0.66\,\mathrm{MPa}$)
of MWNTs has also been estimated by Cumings and Zettl \cite{00-Cumings} in
their discovery of reversible telescopic extensions. The intertube shear
strength of CNT-ropes is of the same order or even lower than the intershell
shear strengths of MWNTs. Although the above-mentioned findings have
inspired many applications including ultra-low friction bearing \cite%
{00-Cumings} and gigahertz oscillators \cite{02-Zheng}, the ultra-low
intershell shear strength allows only negligible load to be transferred from
the outermost shell of MWNTs to the inner ones in spite of the fact that a
typical MWNT contains dozens of walls. This understanding reveals a crucial
mechanism that prevents the creation of CNT-reinforced composites with super
strength.

Approaches have been progressively proposed for improving the intershell or
intertube load-transfer properties of CNTs, which can be categorized
according to two different mechanisms. The basic secrets of spun-yarn
mechanism were discovered long ago and indicated by archaeological evidences
from the late Stone Age. That mechanism was realized by Jiang \emph{et al.} 
\cite{02-Jiang} in the creation of macroscopic self-assembled yarns, which
were drawn out and spun directly from superaligned arrays of CNTs. The
molecular mechanics simulations by Qian \textit{et al.} \cite{03-Qian} on a
CNT rope comprising seven close-packed SWNTs predicted that a remarkable
enhancement of the intertube load transfer property could be achieved by
twisting the rope. The subsequently measured tensile specific strength ($%
0.31 $ to $0.58\,\mathrm{GPa}$) of the spun MWNT yarns by Zhang \emph{et al.}%
\thinspace \cite{04-Zhang}, however, turn out not as much enhanced as
expected from the mechansim put forward by Qian \cite{03-Qian}, and the
Young's modulus of the spun yarns of CNTs may reduce to lower than one-tenth
of that of straight CNTs \cite{04-Zhang, 02-Zhu}. The second mechanism can
be named as crosslink. Kis \emph{et al.}\thinspace \cite{04-Kis} achieved a
breakthrough by using moderate electron-beam irradiation inside a
transmission electronic microscopy to generate crosslinks between the tubes,
effectively eliminating sliding between the nanotubes and leading to a
30-fold increase of the bending modulus. In the light of abinitio total
energy density functional theory, da Silva \textit{et al.} \cite{05-Silva}
showed that Wigner defects existing in SWNT bundles could form a strong link
between nanotubes and increase the shear modulus by a sizable amount.
Huhtala \emph{et al.\thinspace }\cite{04-Huhtala} suggested using small-dose
electron or ion irradiation to partially transfer the load to the inner
shell of double-walled carbon nanotubes, and their molecular-dynamics
simulations showed that a small number of defects could increase the
interlayer shear strength by several orders of magnitude. On the other hand,
however, the coexisting vacancy defects and holes induced during irradiation
may significantly (by 30\% to 90\%) reduce the tensile strength down to the
same level as that of existing commercial carbon fibers \cite{04-Mielke,
04-Sammal}. Furthermore, it was found that interstitial-vacancy pairs (or
Wigner defects) as crosslinks were thermally instable - they will disappear
at temperatures around just 500 K \cite{05-Urita}. Nevertheless, all methods
proposed by now can not enhance the interwall or intertube shear strengths
by several orders of magnitude without remarkably reducing the tensile
strength and/or tensile stiffness.

Stimulated by the above-mentioned observations, we propose a new type of
CNTs called condensed multiwalled carbon nanotubes (CMWNTs), which are
composed of ideal or defect-free coaxial cylindrical shells of sp$^{2}$%
-bonded carbon atoms with interwall spacings smaller than those (about $%
0.34\,\mathrm{nm}$) observed in ordinary MWNTs. Since sp$^{2}$-bond is
physically valid for interwall spacings lowered down to $0.16$\symbol{126}$%
0.18\,\mathrm{nm}$ \cite{02-Brenner}. our idea of CMWNTs is illuminated by
the validity of thin-shell model of SWNTs \cite{96-Yakobson,05-Wang}, in
which the representative thickness about $0.066\,\mathrm{nm}$ was much
smaller than the mean interwall spacing $0.34\,\mathrm{nm}$, and which was
also valid for modeling various complex mechanical behaviors.

Investigations on a six-walled zigzag CMWNTs -
(5,0)@(11,0)@(17,0)@(23,0)@(29,0)@(34,0) were first conducted using the
recently proposed second-generation reactive empirical bond order (REBO)
potential \cite{02-Brenner}, in which both the covalent bonding and van der
Waals interactions into account. We modeled a supercell of the CMWNT
containing 30 lattice loops in the axial directioin or having an axial
length of about $4.26$ nm. Our calculations showed that the energy-optimized
configuration of the CMWNT, under the constraint that the six constituent
shells are consistently elongated or shortened, has indeed compressed
interwall spacings of $0.248$, $0.256$, $0.263$, $0.272$, $0.283\,\mathrm{nm}
$ successively from the innermost shell (Shell 1) to the outermost one
(Shell 6) as depicted in Fig. 1. To estimate the intershell shear strengths,
we axially slided the $i$th shell relatively to the others and then
calculated the energy fluctuation $W_{i}$ versus the slide distance $x$. The
force fluctuation $F_{i}$ is valued as $-dW_{i}/dx$ and the force amplitude $%
F_{\max }^{i}$ should be balanced by the intershell shear strengths as
follows: $F_{\max }^{i}=\tau _{i-1,i}A_{i-1,i}+\tau _{i,i+1}A_{i,i+1}$,
where $\tau _{j,j+1}$ and $A_{j,j+1}$ denote the intershell shear strength
and the middle surface area between the $j$th and $(j+1)$th shells with the
conventions $\tau _{0,1}=\tau _{6,7}=0$. The estimated values of the five
intershell shear strengths are plotted in Fig. 1. For comparison, the
intershell shear strength of the ordinary zigzag double-walled carbon
nanotube (DWNT) - (10,0)@(19,0) - with the interwall spacing of $0.353\,%
\mathrm{nm}$, is estimated to be $0.043\,\mathrm{GPa}$, which is consistent
with other theoretical predictions \cite{03-Guo}. The results indicate that
the intershell shear strength of the investigated CMWNT are in fact,
enhanced greatly, by $40$ to$\ 380$ times. Interestingly, the simulated
intershell shear strengths $\tau (s)$ exponentially depend upon the
interwall spacings $s$ as follows $\tau (s)/\tau (s_{0})=\exp (21.72\frac{%
s_{0}-s}{s_{0}})$, where $\tau (s_{0})=0.0453\,\mathrm{GPa}$ and $%
s_{0}=0.34\,\mathrm{nm}$. This observation suggests a possibility of
tremendous enhancement achieved by further compressing the interwall spacing.

\begin{figure}
\begin{center}
\includegraphics[scale=0.5,angle=270]{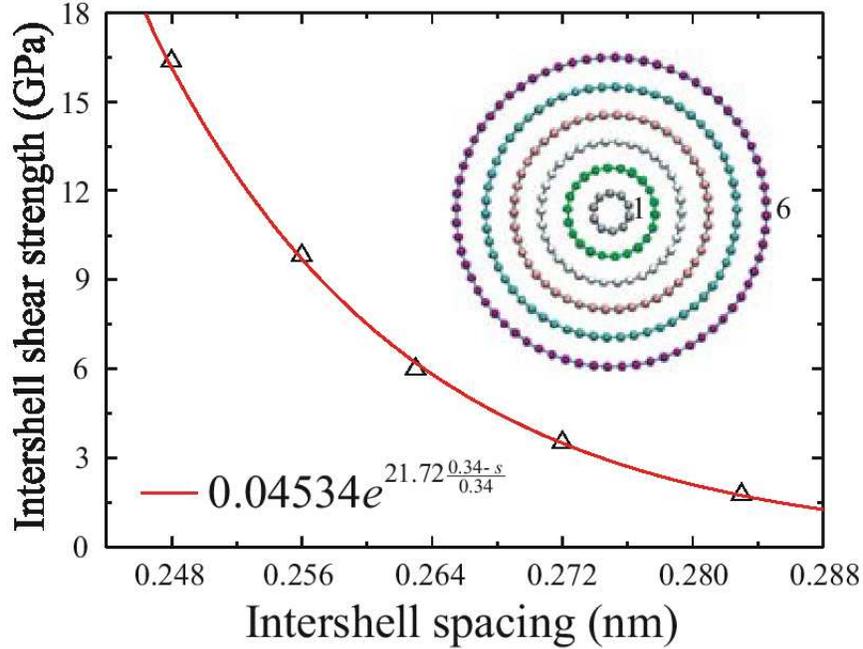}
\end{center}
\caption{Simulated intershell shear
strength versus the compressed interwall spacing (triangles) following an
exponential dependence (fitted line). Insert: the energy-optimized
configuration of the CMWNTs}
\label{figure1}
\end{figure}

As mentioned previously, excellent tensile strength, superior tensile
stiffness, and proper thermal stability are considered a must for the CMWNTs
to become superfibers. To analyze those properties of the investigated
CMWNT, we calculated the pulling force when elongating the surpercell and
then evaluate the nominal stress $\sigma $ defined as the pulling force
divided by the representative cross-section area: $A=\frac{\pi }{4}(D_{6}+%
\bar{s})^{2}$, where $D_{6}$ ($=3.0$ nm) is the diameter of the outermost
shell (34,0) and $\bar{s}$ ($=0.2644\,\mathrm{nm}$) the average interwall
spacing. The calculated $\sigma $ versus the tensile strain or relative
elongation $\epsilon $ for the investigated CMWNT is plotted in Fig. 2. For
comparion, the $\sigma -\epsilon $ curves of the SWNTs (5,0), (11,0), ...,
(34,0) with the assigned representative wall thickness $0.34\,\mathrm{nm}$
are also calculated and illustrated in Fig. 2, in good agreement with those
reported by \cite{04-Mielke}. We note that the CMWNT has a failure strain ($%
\epsilon _{\mathrm{cr}}=18\%$) almost the same as those of the SWNTs while
its failure stress or tensile strength is even 20\% higher than those of the
SWNTs. This observation is exciting, in contrast to the existing methods or
techniques for enhancing intershell shear strengths, which always suffer too
much tensile strength loss. Furthermore, the Young's modulus $Y$ valued as
the slope of the $\sigma -\epsilon $ curve at $\epsilon =0$, namely $Y=\frac{%
d\sigma }{d\epsilon }|_{\epsilon =0}$, of the CMWNT is found larger than
those of the SWNTs by $21.8\%$. Since the basal Young's modulus of graphite
and the axial tensile modulus of SWNTs are approximately $1\,\mathrm{TPa}$,
the largest Young's modulus of all existing materials, our proposed CMWNT
proves to be an uttermost material in respect of both tensile modulus and
tensile strength.

\begin{figure}
\begin{center}
\includegraphics[scale=0.5,angle=270]{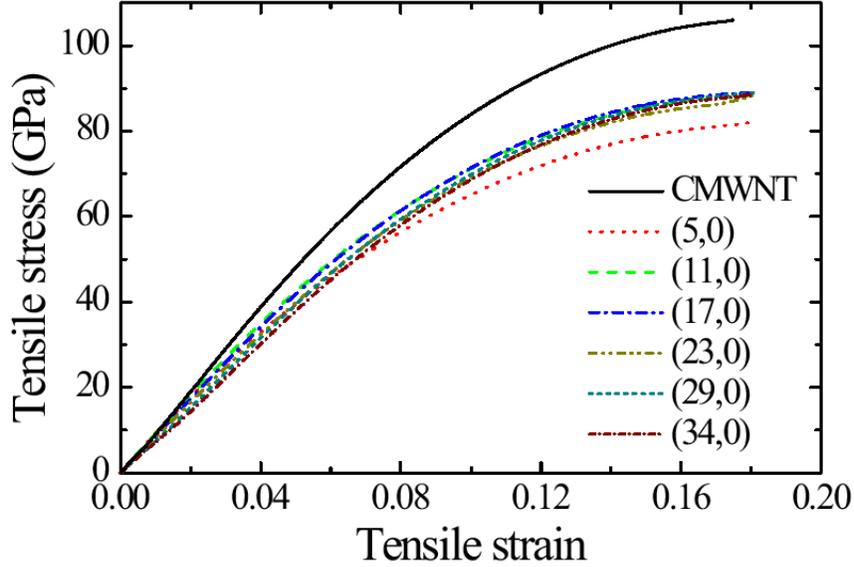}
\end{center}
\caption{$\protect\sigma -\protect%
\epsilon $ curves of the investigated CMWNT and its six constituent shells
as individual SWNTs.}
\label{figure2}
\end{figure}

To understand the mechanism of intershell-shear-strength enhancement, we
modeled the constituent shells of the investigated CMWNT as isotropic
elastic thin-shells \cite{96-Yakobson,05-Wang}. Comparing the diameters of
the constituent shells in the optimized configuration of the investigated
CMWNT with those of these shells in their respective SWNT states, we found
that these shells sustain the circumferential self-equilibrum strains $%
\varepsilon $ of -8.13\%, -4.65\%, 2.69\%, 5.13\%, 7.33\%, and 9.64\%,
respectively, all below the failure strain limit (about $18\%$) range.
Notably the innermost two shells are circumferentially compressed and others
extended. As a rough estimate, the circumferential membranous force per unit
axial length sustained by the $i$th shell is $N_{i}=Yt\varepsilon _{i}$,
which is balanced by the radial pressure difference in the form $%
2N_{i}=p_{i-1,i}D_{i-1,i}-p_{i,i+1}D_{i,i+1}$, where $Yt=0.34\,\,\mathrm{TPa}%
\times \mathrm{\mathrm{nm}}$ is the tensile rigidity, and $p_{j,j+1}$ and $%
D_{j,j+1}$ respectively the interwall pressure and middle surface diameter
between the $j$th and $(j+1)$th shells as schematically illustrated in Fig.
3(a). With the radially free-constraint condition $p_{6,7}=0$ the interwall
pressures $p_{j,j+1}$ for $j=1,2,...,5$ are thus estimated respectively as $%
173.3$, $121.98$, $73.08$, $40.15$, and $17.66\,\mathrm{GPa}$, approximately
exponentially dependent to the compressed interwall spacings. Our direct
atomic calculations using the van der Waals forces between interwall carbon
atoms result in interwall pressures of $113.6$, $78.16$, $52.50$, $33.54$,
and $17.82\,\mathrm{GPa}$ respectively. The existence of these very high
interwall pressures is very likely the primary mechanism of enhancing the
intershell shear strength. Particularly, the following two phenomena which
appeared in the above-mentioned observations interested us. First, the
existence of pressure of the order of magnitude $100\,\mathrm{GPa}$ between
two adjacent core shells implies that CMWNTs may not stably exist without a
very small core. To ascertain this point, we study the two- to five-walled
tubes generated from the investigated CMWNT by drawing out different numbers
of the inner shells, and showed their energy-optimized configurations in
Fig. 3(b) . The collapsed configurations for the two- to four-walled tubes
only yield insignificantly enhancing effects on the interwall shell strength
compared with those of ordinary MWNTs. Second, it is known that a pressure
as high as several to dozens GPa may lead to a sp$^{2}$-sp$^{3}$ transition 
\cite{03-Mao}. It is natural to question the thermal stability of CMWNTs as
highly strained structures. To analyze this problem we performed atomistic
simulations under canonical ensemble with various temperatures. As a result,
the investigated CMWNT as a strained sp$^{2}$-bonded structure remains
stable up to $2,000\,\mathrm{K}$ within hundreds of picoseconds. Above this
critical temperature the sp$^{2}$-bonded innermost shell (5,0) was
transformed into sp$^{3}$-bonded or diamond nanowire; and above $3,000\,%
\mathrm{K}$ the two-walled core (5,0)@(11,0) was transformed into diamond
nanowire. It is found through further observations that the annealed diamond
nanowires produced this way are stable. This study exhibits a new method for
producing diamond nanowires without an extra-high external pressure.

\begin{figure}
\begin{center}
\includegraphics[scale=0.8,angle=0]{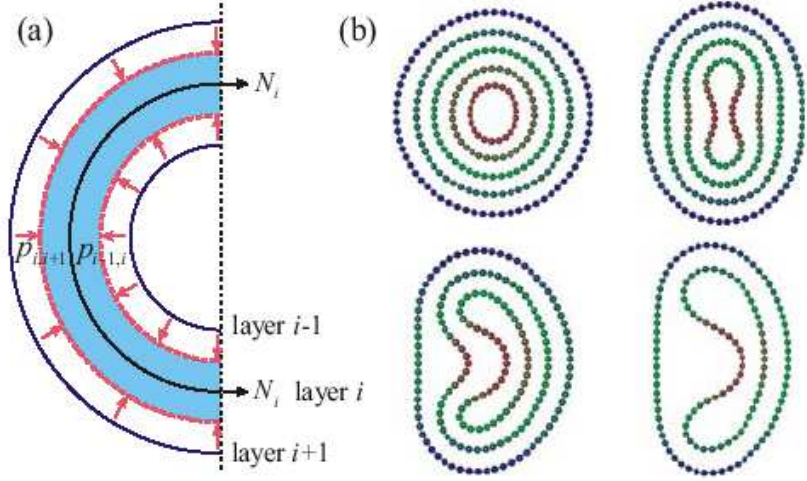}
\end{center}
\caption{Enhancement mechanism and
instability. (a) Balance of circumferential force $N_{i}$ and interwall
pressures $p_{i-1,i}$ and $p_{i,i+1}$; (b) Energy-optimized Configurations
of the five- to two-walled tubes generated by drawing out the inner one to
four shells from the investigated CMWNT.}
\label{figure3}
\end{figure}

The measurements of slide frictions between two relatively rotated
contacting layers of graphite \cite{04-Dienwiebel} show obvious dependence
on the rotation angle, with a remarkable peak corresponding to zigzag
commensuration. Similar observations were reported for the great
commensurating dependence of intershell shear strength of DWNTs \cite{03-Guo}%
. In order to compare the enhancement effects of commensuration, we
investigated the zigzag double-walled tubes (10,0)@(16,0), (10,0)@(17,0),
(10,0)@(18,0), and (10,0)@(19,0), as well as chiral commensurated ones
(11,2)@(10,10), (11,2)@(11,11), and (11,2)@(12,12). Our calculated results
of intershell shear strength versus intershell spacing are presented in Fig.
4, where the solid lines are fitted exponential dependence laws $\tau
(s)=0.0421\exp (53.79(0.353-s))\,\mathrm{GPa}$ for the zigzag tubes and $%
\tau (s)=0.000469\exp (95.08(0.341-s))\,\mathrm{GPa}$ for the chiral ones
respectively. The former is quite closed to the exponential law of the
investigated six-walled zigzag CMWNT. This observation indicates that the
exponential law is approximately independent of the number of walls. More
interestingly, we find that the enhancing effect of spacing-compression on
the chiral tubes is much larger than those of the zigzag ones. Although the
intershell strength of the ordinary zigzag DWNT (10,0)@(19,0) differs from
that of the chiral one (11,2)@(12,12) by two orders, it is expected that the
intershell strengths of the condensed zigzag and chiral DWNTs with the
diameter about $0.22\,\mathrm{nm}$ would become almost the same. In other
words, the chirality effect on enhancement of intershell strength of CMWNTs
appears to be insignificant.

\begin{figure}
\begin{center}
\includegraphics[scale=0.5,angle=270]{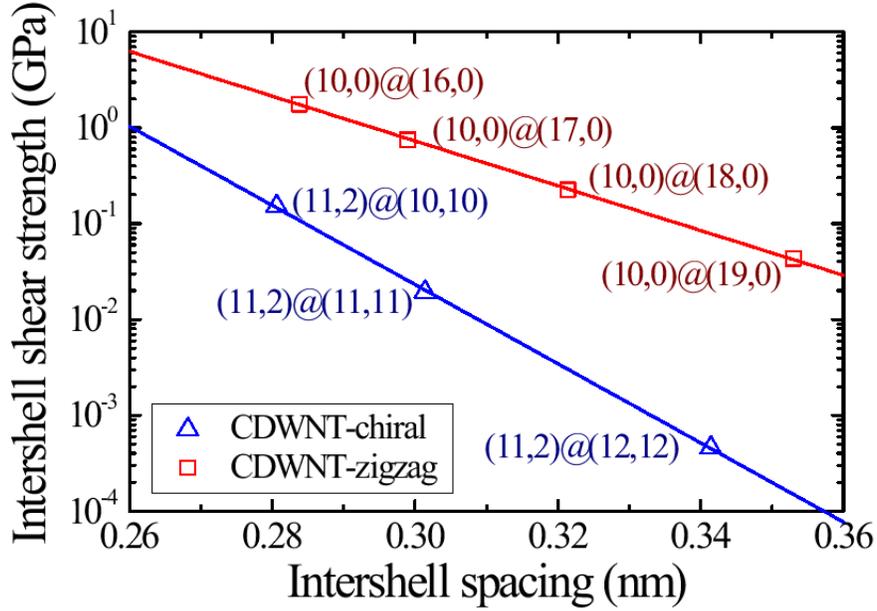}
\end{center}
\caption{Chirality effect on enhancement
of intershell shear strnghth. The enhancement of intershell strength for the
four zigzag CDWNTs and that for the three chiral ones follow different
exponential laws.}
\label{figure4}
\end{figure}

Finally, we would conclude this report by proposing a method of producing
CMWNTs, called irradiation-reconstruction process. Using electron
irradiation within certain temperature circumstances ($\sim $ 600 $^{\circ }$
C), Banhart \emph{et al.} \cite{96-Banhart} found that ordinary carbon
onions as assmeblies of concentric spherical shells of sp$^{2}$-bonded
carbons with interwall spacings about $0.34\,\mathrm{nm}$, could be
transformed into condensed carbon onions with the observed interwall
spacings of $0.22$ to $0.28\,\mathrm{nm}$. The compressing mechanism can be
explained as the following that electron-beam irradiation can knock off
atoms of carbon onions in equilibrium state, thus inducing defects and
vacancies. Since entirely sp$^{2}$-bonded is most energy-favored, anneal
reconstruction could circumferentially shrink the constituent shells and
generate very high intershell pressure because of the nanoscale shell
diameters. Our further molecular dynamics calculations have confirmed that
this mechanism is also valid for producing CMWNTs, indicating that CMWNTs
could be low-costly produced. Therefore, CMWNTs could not only serve as the
excellent candidates for manufacturing space elevators but also promise wide
applications in producing super-reinforced composites.


\begin{thebibliography}{99}
\bibitem{05-Edwards} B. C. Edwards, \textit{IEEE Spectrum} \textbf{42}, 36 (2005); B. I. Yakobson and R. E. Smalley, \textit{Am. Sci.} \textbf{85},324 (1997).
\bibitem{91-Iijima} S. Iijima, \textit{Nature} \textbf{354}, 56 (1991).
\bibitem{96-Treacy} M. M. J. Treacy \textit{et al}., \textit{Nature} \textbf{381}, 678 (1996); E. W. Wong \textit{et al}., \textit{Science} \textbf{277}, 1971 (1997).
P. Poncharal \textit{et al}., \textit{Science} \textbf{283}, 1513 (1999).
\bibitem{00-Yu} M. F. Yu \textit{et al}., \textit{Science} \textbf{287}, 637 (2000).
M. F. Yu, \emph{et al}., \textit{J. Phys. Chem. B} \textbf{104}, 8764 (2000).
\bibitem{92-Robertson} D. H. Robertson \emph{et al}., \textit{Phys. Rev. B} \textbf{45}, 12592 (1992).
\bibitem{96-Yakobson} B. I. Yakobson \textit{et al}., \textit{Phys. Rev. Lett.} \textbf{76}, 2511 (1996).
\bibitem{98-Hernandez} E. Hern\'{a}ndez \textit{et al}., \textit{Phys. Rev. Lett.} \textbf{80}, 4502 (1998).
\bibitem{02-Belytschko} T. Belytschko \textit{et al}., \textit{Phys. Rev. B} \textbf{65}, 235430 (2002).
\bibitem{03-Dumitrica} T. Dumitrica \textit{et al}., \textit{J. Chem. Phys.} \textbf{118}, 9485 (2003); \textit{Appl. Phys. Lett.} \textbf{84}, 2775 (2004).
\bibitem{04-Mielke} S. L. Mielke \textit{et al}., \textit{Chem. Phys. Lett.} \textbf{390}, 413 (2004).
\bibitem{00-Cumings} J. Cumings and A. Zettl, \textit{Science} \textbf{289}, 602 (2000).
\bibitem{02-Zheng} Q. S. Zheng and Q. Jiang, \textit{Phys. Rev. Lett.} \textbf{88}, 045503 (2002).
\bibitem{02-Jiang} K. L. Jiang \textit{et al}., \textit{Nature}. \textbf{419}, 801 (2002).
\bibitem{03-Qian} D. Qian \textit{et al}., \textit{Compos. Sci. Technol.} \textbf{63}, 1561 (2003).
\bibitem{04-Zhang} M. Zhang \textit{et al}., \textit{Science} \textbf{306}, 1358 (2004).
\bibitem{02-Zhu} H. W. Zhu \textit{et al}., \textit{Science} \textbf{296}, 884 (2002).
\bibitem{04-Kis} A. Kis \textit{et al}., \textit{Nat. Mater.} \textbf{3}, 153 (2004).
\bibitem{05-Silva} A. J. R. da Silva \textit{et al}., \textit{Nano Lett.} \textbf{5}, 1045 (2005).
\bibitem{04-Huhtala} M. Huhtala \textit{et al}., \textit{Phys. Rev. B} \textbf{70}, 045404 (2004).
\bibitem{04-Sammal} M. Sammalkorpi \textit{et al}., \textit{Phys. Rev. B} \textbf{70}, 245416 (2004).
\bibitem{05-Urita} K. Urita \textit{et al}., \textit{Phys. Rev. Lett.} \textbf{94}, 155502 (2005).
\bibitem{02-Brenner} D. W. Brenner \textit{et al}., \textit{J. Phys.: Condens. Mater.} \textbf{14}, 783 (2002).
\bibitem{03-Guo} W. L. Guo \textit{et al}., \textit{Phys. Rev. Lett.} \textbf{91}, 125501 (2003).
\bibitem{05-Wang} L. F. Wang \textit{et al}., \textit{Phys. Rev. Lett.} \textbf{95}, 105501 (2005).
\bibitem{03-Mao} W.\ L. Mao \textit{et al}., \textit{Science} \textbf{302}, 425 (2003).
\bibitem{04-Dienwiebel} M. Dienwiebel \textit{et al}., \textit{Phys. Rev. Lett.} \textbf{92}, 126101 (2004).
\bibitem{96-Banhart} F. Banhart and P. M. Ajayan, \textit{Nature} \textbf{382}, 433 (1996); \textit{Adv. Mater.} \textbf{9}, 261 (1997).
\end{thebibliography}
\end{document}